\documentclass{aa}
\usepackage{amssymb,amsmath,graphicx,color,microtype}
\usepackage{txfonts}
\usepackage{multicol}
\usepackage{epsf}
\usepackage{epstopdf}
\usepackage{amssymb}
\usepackage{ulem}
\newcommand{\nc}{\newcommand}
\newcommand\be{\begin{equation}}
\newcommand\ee{\end{equation}}

\usepackage{float}
\usepackage{hyperref}

\nc{\e}{{\bf{e}}}
\nc{\kk}{{\bf{k}}}
\nc{\pp}{{\bf{p}}}

\nc{\bfk}{{\bf{k}}}
\nc{\bfx}{{\bf{x}}}
\nc{\bfp}{{\bf{p}}}

\nc{\eH}{{\epsilon_H}}
\nc{\calP}{{\cal P}}
\nc{\im}{{ \mathrm{Im} } }

\begin{document}

\title{Multi-copy axion transfer function and observational implications of effective de Broglie scales}

   \author{Jiashuo Zhang\fnmsep\thanks{joshua.z.0211@gmail.com}
          \inst{1,2}
          \and
          Tom Broadhurst\inst{1,3,4}
          \and
          Jeremy Lim\inst{2,5}
          \and
          Paloma Morilla\inst{3}
          \and 
          Sung Kei Li\inst{2,5}
          }

   \institute{
            Donostia International Physics Center, Paseo Manuel de Lardizabal, 4, San Sebasti\'an, 20018, Spain
        \and 
            Department of Physics, The University of Hong Kong, Pokfulam Road, Hong Kong SAR
        \and
            Department of Physics, University of the Basque Country UPV/EHU, Bilbao, Spain
        \and 
            Ikerbasque, Basque Foundation for Science, Bilbao, Spain
        \and 
            The Hong Kong Institute for Astronomy and Astrophysics, The University of Hong Kong, Pokfulam Road, Hong Kong SAR
             }
    \date{Received ...; accepted ...}

\abstract {} {Ultra-light axions are viable fuzzy or wave-like dark matter ($\psi$DM) candidates generically predicted by the string axiverse paradigm with multiple particle copies, whereas most of the discussions and constraints on $\psi$DM from astronomical observations to date are based on the assumption of a single-particle copy. We aim to bridge this gap by exploring the generic multi-axion scenario motivated in the string axiverse context and investigate its astronomical implications in the linear and non-linear regimes.} {In the linear regime, we performed a linear density perturbation analysis to investigate the $\psi$DM transfer functions and determined the corresponding suppression scale on the large-scale structure (LSS) in the context of the multi-copy axion. In the non-linear regime within individual galaxies, we investigated the superposition of Gaussian random fields from individual particle copies to determine the net lensing and stellar heating signatures. } {In the linear regime, we provide a simplified prescription for obtaining multi-copy axion transfer functions, and we also identify an equivalence among all axion copies owing to the mutual coupling to the gravitational potential. As a result of this equivalence, we argue that the suppression to LSS is governed by an effective mass $m_{\text{eff}}^{-2}=\sum_i w_i m_i^{-2}$, with $\{ w_i\}$ being fractional contributions of different copies to the full cosmic dark matter density. In the non-linear regime within galaxy haloes, we show that similar notions of effective mass, with expressions provided, govern the collective wave interference and hence determine the net stellar heating rates and the substructure-induced spread of James Webb Space Telescope (JWST) transients near critical curves. Distinctive to the multi-copy scenario, the effective mass governing the net surface-density perturbations within galaxy haloes is generically anticipated to vary radially. This spatial variation leads to different spreading scales for micro-lensed transients at different radial positions. This signature might be tested with future JWST lensing observations.} {}

\maketitle

\section{Introduction}

Axions are named after a laundry detergent and were originally proposed to clean the so-called strong CP problem \citep{Weinberg_axion, Wilczek_axion}. They remain one of the most viable cold dark matter (CDM) candidates at the $10^{-6}$eV scale \citep{Preskill1982cy, Abbott1982af, Dine1982ah}, given the stringent laboratory null results for long-anticipated WIMPs at the tera-electron-volt scale \citep{2410.17036}. Ultra-light axions, with a much lower mass of $\sim 10^{-22}$ eV \citep{Widrow1993, Huetal2000}, are also increasingly recognised as viable cold dark matter candidates \citep{lamhuireview, axion_cosmology}, particularly since their existence is generically predicted in the context of the string axiverse \citep{Svrcek2006,Arvanitaki2010, 2110.02964}. 

The first cosmological simulations of ultra-light axions revealed a dense solitonic core at the centre of every halo, corresponding to the ground state \citep{Schive2014} and with a mass that scales with host halo mass as $M_{\text{sol}}\propto m_{b}^{-1} M_{h}^{1/3}$ \citep{Schive2014Soliton}, surrounded by a halo of self-interfering waves, hence the term wave dark matter ($\psi$DM). The flat density profile of the solitonic core contrasts with the problematic density cusp of the standard heavy particle CDM ($p$CDM) and is characterised by the de Broglie wavelength. For dark matter-dominated dwarf spheroidal (dSph) galaxies, this wavelength is simply determined by the observed core radius of $\sim$ 0.5 kpc \citep{Chen2018, Pozo2020, Pozo2024PhRvD}. Ultra-light axions also naturally suppress the formation of small-scale structure below an inherent Jeans scale of about $10^8M_\odot$, thereby avoiding the missing satellite problem that challenges $p$CDM \citep{Moore1999,Klypin1999}. 

In the $\psi$DM context, the inherent wave interference provides uniquely distinctive lensing-based tests, as the Einstein ring becomes highly corrugated on the de Broglie scale. This effect has been shown to be capable of explaining well-known flux anomalies of lensed QSOs and compact radio sources without requiring a finely tuned population of CDM subhaloes \citep{Chan2020, Amruth}. More recently, \citet{TomKeith2024} have demonstrated that this interference pattern can also account for the broad spread and apparent skewness of microlensed stars on the negative-parity side of the critical curves bisecting lensed galaxies. This matches the JWST observations of the Dragon Arc well \citep{Fudamoto2025NatAs}. In contrast, when CDM subhaloes are added, the transients are confined to a narrower band and are skewed in the opposite direction \citep{2304.06064}.

Despite the phenomenological successes related to dwarf galaxy profiles and lensing, the favoured boson mass scale of $\sim 10^{-22}$ eV is argued to be in tension with the mass constraints from Lyman-$\alpha$ forest analyses, which exclude boson masses below $2\times 10^{-20}$eV \citep{RogersPeiris2020} in order to reproduce the observed forest transmission power spectrum. In addition, no clear faint-end turnover has yet been reported in luminosity functions (LFs) measured in deep surveys at $z>1$, ruling out axions lighter than $2.51\times 10^{-22}$eV \citep{Winch2024ApJ}. These apparently strict constraints, however, assume a single boson model\footnote{Mixed $p$CDM+$\psi$DM models have also been investigated in the literature, e.g. in \citet{Marsh2014MNRAS, Winch2024ApJ}, but we do not discuss these models further.}. This raises the question of whether the data may be better explained within the string axiverse, where multiple axion species are naturally expected, with a discrete axion mass spectrum spanning many decades \citep{Svrcek2006, 2103.06812}. Tentative support from astronomical observation for such a multiple-axion scenario comes from the possible nested soliton structure within galaxy cores \citep{1811.03771}, where the common occurrence of nuclear star clusters within galaxy cores might imply a second axion species with mass $\sim 10^{-20}$ eV.

Further insights into the diversity of halo structures predicted in the multi-axion context are now being provided by the first cosmological simulations incorporating two axion species, where sufficient resolution can be achieved for two axion species differing by up to a factor of 5 in boson mass \citep{Luu2024, Pozo2025}. We anticipate that such simulations will be generalised to higher mass ratios and additional axion copies, even though generating the relevant initial density perturbations for each species probably is difficult and time-intensive \citep{Preston2025MNRAS}. 

In Sec.\ref{sec_linearregime} we provide a practical prescription for obtaining the multi-axion transfer functions that can be extended to an arbitrary number of axion copies and relative density contributions based on a linear perturbation analysis. We also demonstrate that the collective behaviour of all copies is well described by an effective boson mass, $m_{\text{eff}}$, arising from mutual gravitational coupling on scales close to the halo Jeans scale. Consequently, the suppression of large-scale structure and the faint-end turnover of the galaxy LF are expected to be characterised simply by this effective boson mass scale.

In Sec.\ref{sec_nonlinearregime} we show that the notion of an effective mass extends to the discussion of collective wave interference on smaller scales within dark matter haloes (i.e. the non-linear regime), thereby determining the corrugation of critical curves in lensing and accounting for flux and positional anomalies of lensed images \citep{Chan2020, Amruth}, as well as the location of transient lensed stars \citep{TomKeith2024}. This effective boson mass ($m_{\text{eff}}'$, not to be confused with $m_{\text{eff}}$ because they are defined in entirely different contexts) depends on local relative axion densities and can be determined by considering the axion species indicated by dwarf galaxies \citep{Pozo2024PhRvD}. The inherent spatial variation in $m_{\text{eff}}'$, reflecting different degrees of radial concentration of the different mass copies, might also provide a distinctive lensing signature testable with future JWST observations.

In the last part of Sec.\ref{sec_nonlinearregime}, we connect our considered multi-copy framework with higher-spin dark matter and comment on the corresponding implications. We also comment on the prospects for stellar heating, but leave a more detailed investigation for future work. We also leave the effect of non-trivial particle potential (e.g., in \citet{2112.09337}) for future investigation, and consider only free axion fields without self-interaction or cross-mixing throughout this work. Finally, a concise summary of the results is given in Sec.\ref{sec_summary}.

\section{Multi-copy axion linear perturbation equations}
\label{sec_linearregime}

In this section, we consider the linear perturbation regime and the multi-axion transfer function relevant for determining the initial density perturbations of different axion species in future cosmological simulations. For simplicity, we assumed all copies to have coherent phases in matter density and absence of isocurvature modes, that is, axions have homogeneous relative fractions to dark matter. We began with the Schr\"{o}dinger-Poisson equation for an arbitrary number of axion fields $\varphi_i$ in a FLRW expanding Universe,

\begin{equation}
    i(\partial_t + \frac{3}{2}\mathcal{H} )\varphi_i = -\frac{\nabla^2 \varphi_i}{2m_ia^2} + \Phi m_i \varphi_i ,
\end{equation}

\begin{equation}
    \nabla^2 \Phi = 4\pi G a^2 \sum_i \delta \rho_i .
\end{equation}
Here, $a$ is the scale factor, $\mathcal{H}\equiv \dot{a}/a$ is the Hubble rate, $\dot{} \equiv d/dt$, $\rho_i = m_i^2|\varphi_i|^2$, and $\delta\rho_i$ describes the perturbation for each copy. To derive the perturbation equations, we performed a Madelung decomposition with $\varphi_i = \sqrt{\rho_i} e^{iS_i}/m_i$ and rewrote the Schr\"{o}dinger-Poisson system as 

\begin{align}
\dot{\rho_i} + 3\mathcal{H} \rho_i + \frac{1}{a} \nabla\cdot (\rho_i v_i) = 0 \\
\dot{v_i} + \mathcal{H} v_i+ \frac{1}{a^3} \nabla Q_i + \frac{1 }{ a}(v_i \cdot \nabla)v_i + \frac{1}{a} \nabla \Phi = 0,
\end{align}
where the velocity field is defined as $v_i \equiv \nabla S_i/(m_i a)$. $Q_i$ denotes the quantum pressure for each axion copy and was defined as

\begin{equation}
    Q_i \equiv -\frac{\nabla^2(\sqrt{\rho_i}) }{2m_i^2 \sqrt{\rho_i}} = -\frac{1}{2m_i^2}\bigg{(} \frac{\nabla^2 \rho_i}{2\rho_i} + \frac{(\nabla \rho_i)^2}{4\rho_i^2} \bigg{)}.
\end{equation} 

By introducing the density contrast $\delta_i \equiv \delta \rho_i/\overline{\rho_i}$, we derived a set of linear perturbation equations to first order in $\delta_i, v_i$. These were collectively expressed in matrix form as

\begin{strip}
\begin{equation}
    \partial_t^2 \begin{pmatrix}
    \delta_1\\
    \vdots \\
    \delta_i \\
    \vdots \\
    \delta_N
\end{pmatrix} + 2 \mathcal{H} \partial_t \begin{pmatrix}
    \delta_1\\
    \vdots \\
    \delta_i \\
    \vdots \\
    \delta_N
\end{pmatrix} - \begin{pmatrix}
    4\pi G\overline{\rho_1}-\frac{k^4}{4m_1^2 a^4}   & \cdots & 4\pi G\overline{\rho_j}  & \cdots & 4\pi G\overline{\rho_N} \\
    4\pi G\overline{\rho_1}   & \cdots & 4\pi G\overline{\rho_j}  & \cdots & 4\pi G\overline{\rho_N} \\
    4\pi G\overline{\rho_1} & \cdots & 4\pi G\overline{\rho_j}-\frac{k^4}{4m_j^2 a^4} & \cdots  & 4\pi G\overline{\rho_N} \\ 
    4\pi G\overline{\rho_1}   & \cdots & 4\pi G\overline{\rho_j}  & \cdots & 4\pi G\overline{\rho_N} \\
    4\pi G\overline{\rho_1}   & \cdots & 4\pi G\overline{\rho_j}  & \cdots & 4\pi G\overline{\rho_N} -\frac{k^4}{4m_N^2 a^4}
\end{pmatrix} \begin{pmatrix}
    \delta_1\\
    \vdots \\
    \delta_j \\
    \vdots \\
    \delta_N
\end{pmatrix} = 0.
\label{Matrix_constrast_equation}
\end{equation}
\end{strip}

We can define a total perturbation contrast as $\overline{\rho_{\text{tot}}}\delta_{\text{tot}} = \sum_i\overline{\rho_i} \delta_i $, then the time derivatives of $\delta_{\text{tot}}$ satisfy

\begin{equation}
    \dot{\delta_{\text{tot}}} = \frac{\sum_i\overline{\rho_i} \dot{\delta_i} }{\overline{\rho_{\text{tot}}}}, \;\;  \ddot{\delta_{\text{tot}}} = \frac{\sum_i\overline{\rho_i} \ddot{\delta_i} }{\overline{\rho_{\text{tot}}}}. 
\label{total_contrast_def}
\end{equation} 
Summing the density contrast equations for all axion copies, we obtained

\begin{equation}
    \ddot{\delta_{\text{tot}}} + 2 \mathcal{H} \dot{\delta_{\text{tot}}} - 4\pi G \delta_{\text{tot}}\overline{\rho_{\text{tot}}}+ \frac{k^4}{4a^4 m_{\text{eff}}^2}\delta_{\text{tot}} = 0,
\label{limiting_case_total}
\end{equation}  
where the effective mass scale is defined by 

\begin{equation}
    \frac{1}{m_{\text{eff}}^2} \delta_{\text{tot}} \overline{\rho_{\text{tot}}} \equiv \sum_i \frac{\overline{\rho_i} \delta_i}{m_i^2 }.
\label{effective_mass_defin}
\end{equation} 
\noindent We assumed the adiabatic growing modes to dominate for all $\delta_i$ and that they evolve proportionally.

\begin{figure*}[h!]
    \centering
    \includegraphics[width=\linewidth]{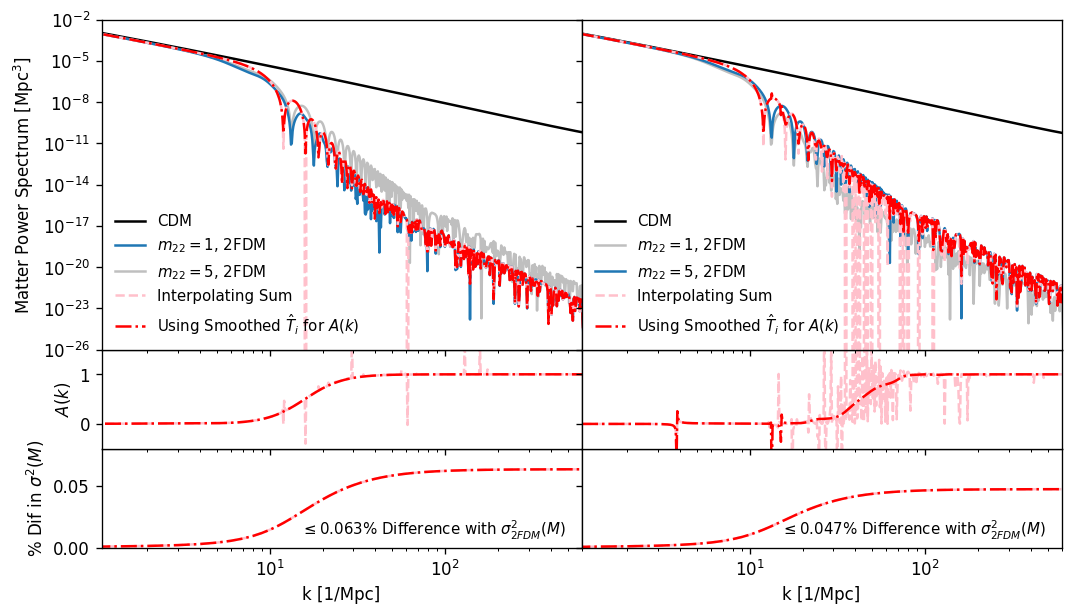}
    \caption{ Testing the validity of interpolating sum ansatz presented by Equ.\ref{density_decomposition} and Equ.\ref{transferfunc_decom} against the double-copy (2FDM) power spectrum from \citet{Luu2024} with full integration of 2FDM perturbation equations. Dark matter is taken to consist of 30\% $m_{22}=1$ and 70\% $m_{22}=5$ axions. In the top panels, the solid blue line presents a 2FDM power spectrum of either $m_{22}=1,5$ copy to be compared with our interpolating sum ansatz (dashed pink and dash-dotted red lines), and the solid light grey line presents the 2FDM power spectrum of the companion copy. The middle panels present the corresponding $A(k)$ factor for full single-axion transfer functions (pink line) or for the envelope function of $\hat{T}_i$s alone, obtained through smoothed spline fitting (red line). The bottom panels then compare the percentage difference in $\sigma^2(M(R=\frac{2\pi}{k}))$ with the 2FDM power spectrum.}
    \label{compare_2FDM}
\end{figure*}

Eq. \eqref{limiting_case_total} is essentially identical to the single-axion case. Hence, the corresponding transfer function\footnote{Here, transfer functions are defined relative to the primordial power spectrum, not relative to particle CDM transfer functions, as done in \citet{Huetal2000, 2201.10238}. Furthermore, to avoid confusion, we will use hatted notation for transfer functions in the single-axion scenario, while unhatted notation is reserved for the multi-copy context.} for $\delta_{\text{tot}}$ can be well approximated by the solution first introduced by \citet{Huetal2000},

\begin{equation}
    \hat{T}_{\text{tot}} \approx \bigg{|}\frac{\cos x^3}{1+x^8}\bigg{|}T_{pCDM}, \; \; x = \frac{1.61}{9}\frac{k}{m^{4/9}_{\text{eff}}},
\label{transfer_1}
\end{equation} 
in which the mass $m_{\text{eff}}$ is in units of $10^{-22}$ eV, the wavenumber $k$ is units of Mpc$^{-1}$, $x$ is an dimensionless quantity (in the discussion below, quantities with undefined units are implicitly dimensionless), and $T_{pCDM}$ is the transfer function of $p$CDM without any wave-induced small-scale suppression. An alternative fitting form with effective fluid formations was later provided by \citet{2201.10238},

\begin{equation}
    \hat{T}_{\text{tot}} \approx \bigg{|}\frac{\sin x^{5/2}}{x^{5/2}(1+Bx^{7/2})}\bigg{|}T_{pCDM}, \; \; x = \frac{A}{9}\frac{k}{m^{1/2}_{\text{eff}}},
\label{transfer_2}
\end{equation} 
where $A = 2.22 m_{\text{eff}}^{1/25 - \ln(m_{\text{eff}})/1000}$, $B = 0.16 m_{\text{eff}}^{-1/20}$.

Likewise, transfer functions computed with the axion-included version of CAMB \citep{CAMB2000ApJ} Boltzmann codes (e.g. AxionCAMB, \citet{axioncamb}; AxiECAMB, \citet{2412.15192}) for the single-axion case might also be applied.

The two limiting cases discussed above can be described by the following equations for each axion copy with mass $m_i$:

\begin{align}
    \ddot{\hat{\delta}}_i + 2 \mathcal{H} \dot{\hat{\delta}}_i - 4\pi G \overline{\rho_{\text{tot}}} \hat{\delta}_i + \frac{k^4 \hat{\delta}_i }{4a^4m_{I}^2} = 0  ,
\label{limiting_cases_1} \\
    \ddot{\hat{\delta}}_{\text{tot}} + 2 \mathcal{H} \dot{\hat{\delta}}_{\text{tot}} - 4\pi G \overline{\rho_{\text{tot}}} \hat{\delta}_{\text{tot}} + \frac{k^4 \hat{\delta}_{\text{tot}} }{4a^4m_{\text{eff}}^2} = 0,
\label{limiting_cases_2},
\end{align}  
where $\hat{\delta}_{\text{tot}}$ and $\hat{\delta}_i$ denote the single-axion solutions for $\delta_i$ under the two limiting cases. In the $\hat{\delta}_i$ equation, the gravitational term should in principle be $4 \pi G \overline{\rho_i} \hat{\delta}_i$. However, since this regime corresponds to the high $k$ limit, where the gravity is subdominant, $\overline{\rho_i}$ can be replaced with $\overline{\rho_{\text{tot}}}$.

We posit that the transition between the two limiting cases for each axion copy can be approximated as 
\begin{equation}
    \delta_i = A(k) \hat{\delta}_i + (1-A(k))\hat{\delta}_{\text{tot}},
    \label{density_decomposition}
\end{equation}
where the frequency-dependent coefficient $A(k)$ approaches 0 at low $k$ and $1$ at high $k$. Here, $\delta_i$ satisfies the original perturbation equation
\begin{equation}
    \ddot{\delta}_i + 2 \mathcal{H} \dot{\delta}_i - 4\pi G \overline{\rho_{\text{tot}}} \delta_{\text{tot}} + \frac{k^4 \delta_i }{4a^4m_i^2} = 0,
\end{equation}
as shown by Eq. \eqref{Matrix_constrast_equation}. As $\hat{\delta}_{\text{tot}} = \delta_{\text{tot}}$ by Eq. \eqref{total_contrast_def}, substituting the above decomposition and using the limiting-case equations gives

\begin{equation}
4\pi G \overline{\rho_{\text{tot}}} A(\hat{\delta}_i - \hat{\delta}_{\text{tot}}) = (1-A)\frac{k^4}{4a^4}\bigg{(} \frac{1}{m_{\text{eff}}^2} - \frac{1}{m_i^2} \bigg{)} \hat{\delta}_{\text{tot}}.
\end{equation}
Hence, the solution for $A(k)$ is 
\begin{equation}
    A(k) = \frac{\frac{k^4}{4a^4}\bigg{(} \frac{1}{m_{\text{eff}}^2} - \frac{1}{m_i^2} \bigg{)} \hat{\delta}_{\text{tot}}}{4\pi G \overline{\rho_{\text{tot}}}(\hat{\delta}_i - \hat{\delta}_{\text{tot}})+\frac{k^4}{4a^4}\bigg{(} \frac{1}{m_{\text{eff}}^2} - \frac{1}{m_i^2} \bigg{)} \hat{\delta}_{\text{tot}}}.
\end{equation}

This expression can be simplified using that $\hat{\delta}_{\text{tot}}/\hat{\delta}_i = \hat{T}_{\text{tot}}/\hat{T}_i$ and the definition of comoving quantum Jeans scale $k_{J}$ \citep{2211.01523} for $\delta_{\text{tot}}$ (given $\overline{\rho_{\text{tot}}}$),

\begin{equation}
4\pi G\overline{\rho_{\text{tot}}}\cdot4a^4 \equiv \frac{a^3 k_{J}^4}{m_{\text{eff}}^2} = 6 \mathcal{H}^2 \Omega_{DM}.
\label{jeans_scale}
\end{equation}
With this, the solution for $A(k)$ reads

\begin{equation}
    A(k) = \frac{\frac{k^4}{\tilde{k}^4}\bigg{(} 1 - \frac{m_{\text{eff}}^2}{m_i^2} \bigg{)}}{\frac{\hat{T}_i}{\hat{T}_{\text{tot}}} -1+\frac{k^4}{\tilde{k}^4}\bigg{(} 1 - \frac{m_{\text{eff}}^2}{m_i^2} \bigg{)}}
\label{Ak_solution_simplified},
\end{equation}
\noindent where

\begin{equation}
    \tilde{k} \equiv a^{3/4}k_{J} \approx \frac{44.7}{\text{Mpc}}\,\bigg{(}6E(a)\frac{\Omega_{_{\text{m}0}}}{0.3}\bigg{)}^{1/4} \bigg{(}\frac{H_0}{70 \,\frac{\text{km/s}}{\text{Mpc}}}\frac{m_{\text{eff}}}{10^{-22} \, \text{eV}}\bigg{)}^{1/2},
\end{equation}
and $E(a) \equiv a^4(\Omega_{\Lambda0}+\Omega_{\text{m}0}/a^3)$.

The transfer function for each axion copy then takes the form 

\begin{equation}
    T_i = A \hat{T}_i + (1-A) \hat{T}_{\text{tot}}, 
    \label{transferfunc_decom}
\end{equation}
where $\hat{T}_i$ and $\hat{T}_{\text{tot}}$ are the single-axion transfer functions. The total transfer function is given by $T_{tot} = \sum_i \overline{\rho_i}T_i /\overline{\rho}_{tot}$. We test the accuracy of the above simple ansatz against the double-copy axion (2FDM) integrator of \cite{Luu2024} in Fig.\ref{compare_2FDM}, assuming dark matter is composed of 30\% $1\times10^{-22}$eV axion and 70\% $5\times10^{-22}$eV axion. Our simple ansatz captures the overall suppression in the power spectrum by $\psi$DM well, except the oscillatory features, which are dominated by the heavier $m_{22}=5$ copy. This dominance arises because the earlier clustering of heavier copy creates a local gravitational potential for lighter copy to follow, breaking our assumption of coherent phases in matter densities. In terms of halo formation, however, we verified that the inability to reproduce the oscillatory features has only a minimal effect, leading to a difference of $\lesssim 0.047\%$ and $\lesssim 0.063\%$ in $\sigma^2(M)$ for the $m_{22}=1$ and $m_{22}=5$ copy at the smallest scale at most. 

\begin{figure}
    \centering
    \includegraphics[width=1\linewidth]{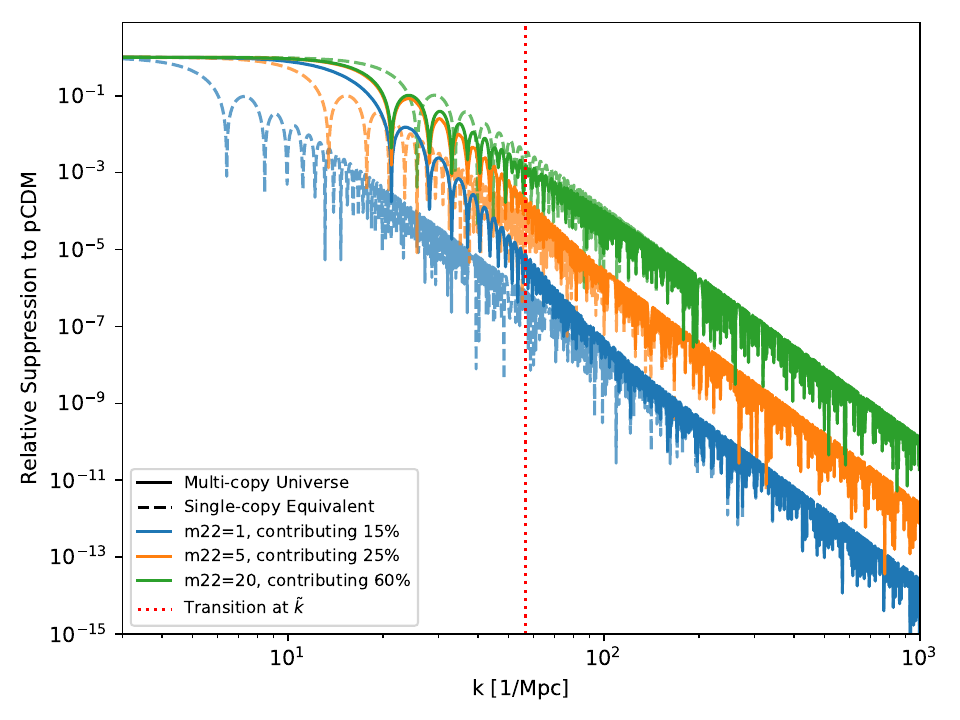}
    \caption{Relative suppression to $p$CDM (solid lines) for each axion copy at redshift $z=127$. The three copies have masses $1, 5$, and $20\times10^{-22}$ eV, accounting for $15\%, 25\%$, and $60\%$ of the total dark matter density $\overline{\rho_{\text{tot}}}$, respectively. The relative suppression for single-copy axion Universe are based on Eq. \eqref{transfer_2}. We also indicate the scale of $\tilde{k}$ using a vertical dotted red line. At $k\gg \tilde{k}$, the axion copies asymptote to their respective single-copy Universe solution (dashed lines in the same respective colour). }
    \label{example_3_copy}
\end{figure}

We emphasise that the power of our simple interpolating sum ansatz is that it can be easily extended to more axion copies. Fig.\ref{example_3_copy} illustrates an example at redshift $z=127$ for a universe with three axion copies of masses $1, 5$, and $ 20\times10^{-22}$eV, contributing $15\%, 25\%$, and $60\%$ of the total matter density $\overline{\rho_{\text{tot}}}$, respectively.

Finally, in Eq. \eqref{transfer_1}, the oscillatory  $\cos x^3$ (or $\sin x^{5/2}$ in Eq. \eqref{transfer_2}) factor can be regarded as a higher-order effect in the determination of $A(k)$. In practice, Eq. \eqref{Ak_solution_simplified} might therefore be implemented by using the ratio of the envelope functions of the transfer functions, for instance $\hat{T}_a /\hat{T}_b = (1+x_b^8)/(1+x_a^8)$ for Eq. \eqref{transfer_1} (and analogously for Eq. \eqref{transfer_2}). For transfer functions obtained from Boltzmann codes, smoothing procedures might be applied to extract the corresponding envelopes. For instance Fig.\ref{compare_2FDM}, shows that such a smoothing exercise has only minimal effects on the final multi-copy power spectrum and no effects on $\sigma^2(M)$. 

\section{Effective boson mass for haloes in the non-linear regime}
\label{sec_nonlinearregime}

We next studied dark matter haloes, whose dynamics lie firmly in the non-linear regime, with the goal of defining an effective de Broglie scale relevant for interpreting lensing signatures and stellar heating effects of $\psi$DM.

For each axion copy, the interference pattern with characteristic scale $\lambda_{\text{dB}, i} \sim 2\pi\hbar /(m_i\sigma)$ fully modulates the local dark matter density. In gravitational lensing, which depends on the projected potential, line-of-sight integration through the halo averages over many independent perturbations of this same de Broglie scale. By the central limit theorem, the resulting projected perturbation to the surface density $\Sigma$ approximates a Gaussian random field (GRF) \citep{Amruth}. At projected radius $R$, the surface density contrast is given by
\begin{equation}
    \frac{\delta\Sigma_i(R)}{\Sigma_i(R)} \sim \sqrt{\frac{\lambda_{\text{dB}, i}}{R_h}},
\end{equation}
where $R_h$ is the effective halo size at the Einstein radius \citep{Venumadhav2017ApJ,  Kawai2022ApJ}.  This GRF pattern produces percent-level surface-density fluctuations that broaden the critical curves of galaxy and cluster lenses, replacing the sharply peaked curve of a smooth profile with a corrugated band \citep{Chan2020, Amruth}. This feature has recently been invoked to explain the $\sim$4 kpc spread of transients observed along the Dragon Arc critical curve \citep{TomKeith2024}, consistent with a de Broglie scale of around 10 pc.

More generally, this scale can be interpreted as an effective de Broglie scale arising from the collective interference of multiple axion copies. On small scales, the copies decouple, so the perturbations behave like a weighted sum of independent GRFs, each distributed as $\mathcal{N}(\Sigma_i,\delta\Sigma_i)$ with weights proportional to the local surface-density contributions. Treating each copy as an independent GRF is justified by the fact that $\delta\Sigma_i$s are only parametrising the statistic of density perturbation along a line of integration independent of the correlation of the phases in the axion fields. \citet{2301.07114} also demonstrated by explicitly calculating two-point correlation functions that multiple copies (with a difference of a small factor of 2 in mass or with equal masses) do not become strongly correlated inside the same halo. For larger differences in the mass factor $(>10)$, as often anticipated by the string axiverse, phase correlations are even less likely to be developed. The net GRF for surface-density fluctuations then has a mean $\Sigma_{\text{tot}}=\sum_i \Sigma_i$ and a standard deviation
\begin{equation}
    \delta \Sigma_{\text{tot}} = \sqrt{\sum_i \delta \Sigma_i^2} = \sqrt{\sum_i \Sigma_i^2\frac{\lambda_{\text{dB}, i}}{R_h}} \equiv \Sigma_{\text{tot}} \sqrt{\frac{\lambda_{\text{dB, eff}}'}{R_h}},
\end{equation}
where the last identity defines an effective scale. The corresponding effective mass, assuming the same velocity dispersion $\sigma$ for all axion copies, is
\begin{equation}
    \frac{1}{m_{\text{eff}}'}  = \sum_i \bigg{(}\frac{\Sigma_i}{\Sigma_{\text{tot}} }\bigg{)}^2 \frac{1}{m_i},
\end{equation}
which encapsulates the net lensing impact of the collective interference pattern.

The lensing-based effective mass $m_{\text{eff}}'$ can be compared to the independent soliton-based boson mass constraints from dark matter-dominated dwarf galaxies. For dSph galaxies, some analyses suggested a characteristic mass of $10^{-22}$ eV \citep{Schive2014, Chen2018}, while more recently, an additional heavier boson has been found to describe the class of ultra-faint dwarf galaxies \citep{Pozo2024PhRvD}. The comparison of the $m_{\text{eff}}'$ inferred from lensing with these two soliton-based masses provides a way to estimate the relative density contributions of different axion copies.

For the Dragon Arc, the favoured de Broglie scale of 10 pc from the spread of transients corresponds to an effective mass of $m_{\text{eff}}' \sim (2.40 \pm 0.03) \times10^{-22}$ eV using $\lambda_{\text{dB}}$ of the order of the soliton core size, which is $ \sim 15 (\frac{10^{-22}\text{eV}}{m}) (\frac{10^{15}M_\odot}{M_{\text{H}}})^{1/3}$ pc \citep{TomKeith2024}, and adopting a halo mass of $M_{\text{H}}\sim (2.4\pm 0.1) \times 10^{14}M_\odot$ as enclosed by the critical curve (based on the WSLAP+ \cite{1609.04822}, CATS \cite{1611.01513} and CANUCS \cite{2403.07062} models). When we assume that the two axion copies suggested for dwarf galaxies, with a mass of $m_1\sim 1.9\times10^{-22}$eV and $m_2\sim 2.3\times 10^{-21}$eV \citep{Pozo2024PhRvD}, make up the halo of Abell 370, the $m_{\text{eff}}'$ favoured by the spread of transients implies fractional contributions of $w_1' = \Sigma_1/\Sigma_{\text{tot}} = 0.889\pm 0.004 $ and $w_2'= \Sigma_2/\Sigma_{\text{tot}} = 0.111 \pm 0.004$ near the Dragon Arc. These values are consistent with expectations that the lighter copy dominates perturbations in the outer region of the halo, while the heavier copy is concentrated in the central region, as shown by \cite{2309.05694}.

\begin{figure}
    \centering
    \includegraphics[width=\linewidth]{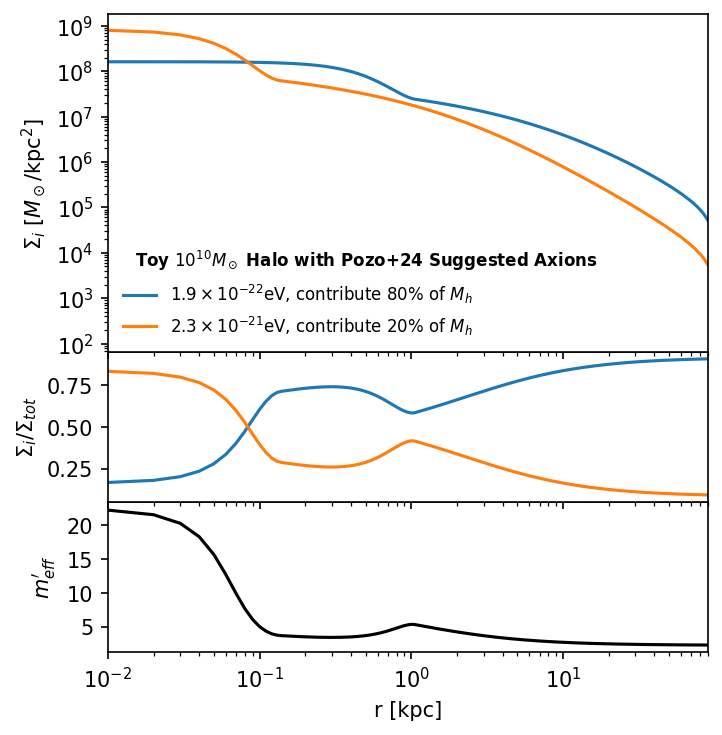}
    \caption{Radial variation in $m_{\text{eff}}'$ with a toy halo of mass $10^{10} M_\odot$ comprising local dwarf galaxy-motivated axions \citep{Pozo2024PhRvD}, where the $1.9\times10^{-22}$eV copy and $2.3\times10^{-21}$eV copy each contribute 80\% and 20\% of the total halo mass. The first panel presents the projected surface density from the respective soliton-NFW profiles, with halo parameters obtained through the estimator of \cite{Liaoetal}. The middle panel presents the relative contribution $w_i'$s to the total surface density. The bottom panel presents the corresponding spatial variation in $m_{\text{eff}}'$, demonstrating a modest radial decrease at $r>1$kpc and stronger variation closer to solitons. As previously mentioned, here the density profiles are taken to be soliton-NFW profiles only suitable for single-axion haloes, whereas the true density profiles of multi-copy haloes may be more complicated (e.g. see \cite{2212.14288, 2309.05694}). Dedicated simulations will be needed to establish the correct density profiles in the multi-copy axion haloes.}
    \label{radial_variation_toy_halo}
\end{figure}

These estimates depend on specific mass choices, while the true axion spectrum in galaxies and clusters is uncertain. Nevertheless, the presence of multiple axion copies may be inferred independently of the detailed mass spectrum if the spread of transients (and thus, $m_{\text{eff}}'$) varies with radial position or source redshift. These dependences trace the spatial variation in the relative contributions $\{ w_i'\}$. This feature is absent in single-copy scenarios. We demonstrate this spatial variation with a toy halo of mass $10^{10}M_\odot$ in Fig.\ref{radial_variation_toy_halo}, in which $m_{\text{eff}}'$ spans a modest range (from $2.3$ to $5.4$) at $r>1$kpc and displays more rapid variation closer to solitons. Similarly, galaxy-gaalaxy strong lensing might provide evidence of multiple copies if the inferred $m_{\text{eff}}'$ varies spatially. An especially promising case is the Cosmic Horseshoe, where the expected $\sim 60$ transients per JWST pointing at 29 AB mag \citep{keith_cosmic_horseshoe} will allow us to determine $m_{\text{eff}}'$ near the tangential critical curve, while the detection or absence of compact sources near radial arcs might constrain the $\psi$DM mass separately \citep{2505.24373}. A discrepancy between these two radial positions (separated by $\sim 5$") would suggest the presence of multiple axion copies. Likewise, a potential signature of multiple axion copies would be a discrepancy between the soliton size inferred from stellar kinematics and the soliton size predicted for the halo given the effective mass $m_{\text{eff}}'$ derived from lensing.

We note that the multi-copy framework can be naturally extended to higher-spin dark matter, such as vector dark matter (VDM; \citealt{2203.11935}), which can be modelled as three independent scalar fields of identical mass. In the VDM scenario, the central soliton size matches that of scalar $\psi$DM, but the density perturbations from collective wave interference are reduced by a factor of $1/\sqrt{3}$. This reduction is also apparent from our expression for $m_{\text{eff}}'$: with three identical copies of mass $m$ and $\Sigma_i/\Sigma_{\text{tot}} = 1/3$, we obtain $m_{\text{eff}}' = 3m$, corresponding to a net perturbation reduced by $1/\sqrt{3}$. For haloes dominated by a single axion, future lensing measurements of $m_{\text{eff}}'$, combined with independent determinations of solitonic core sizes, might therefore help us to distinguish scalar from vector dark matter models. 

Finally, we briefly discuss the implications of multiple copies for stellar heating. As discussed by \cite{Lam2017PhRvD, BarOr2019ApJ, Yang2024MNRAS}, stellar heating by $\psi$DM can be approximated as scattering with quasi-particles of granular mass $M_{\text{gran}}\sim \rho_h \lambda_{\text{dB}}^3$, where $\rho_h$ is the halo density. Through energy conservation, each granule-star encounter can be taken to change the kinetic energy of stars by $v\delta v \sim \frac{GM_{\text{gran}}}{\lambda_{\text{dB}}}$, and hence, the velocity dispersion $\sigma^2$ by $ \sim \delta v^2 \sim \frac{G^2M_{\text{gran}}^2}{\lambda_{\text{dB}}^2 v^2 }$. Together with the encountering rate $\sim v/\lambda_{\text{dB}}$ as $\lambda_{\text{dB}}$ characterises the typical separation of granules, the net local heating rate $\frac{d\sigma^2}{dt}$ then scales as $\sim G^2\rho_h^2\lambda_{\text{dB}}^3 \sim \rho_h^2/m_b^3$. We argue that this scaling relation extends to the multi-copy case with same proportionality coefficients for axion species, including the dropped explicit velocity and velocity dispersion dependence because dark matter particles are anticipated to have similar velocity distributions in the same halo. \cite{BarOr2019ApJ} reported that $\lambda_{\text{dB}}$ also entered the proportionality coefficients through the Coulomb logarithm $\log \Lambda_{\text{FDM}} = \log \frac{2b_{max}}{\lambda_{\text{dB}}}$ characterising the ratio of the maximum (on scales of a halo size or stellar orbit radius) and minimum (granule size) scales of star-granule encounters. Typically for stellar heating, however, we have $\lambda_{\text{dB}} \ll b_{max}$, and hence, $\log \Lambda_{\text{FDM}}$ can be safely taken to be $ \approx \log b_{max}$ without the $\lambda_{\text{dB}}$ dependence. In the multi-copy case, the net change in velocity dispersion $\Delta\sigma$ can be treated as coming from kicks by independent GRFs integrated along the stellar orbits over a timescale of $\Delta t$. The net heating rate then follows from the quadrature sum of the contributions  of each copy, yielding an effective mass $m_{\text{eff}}''$ analogous to that for lensing, 
\begin{equation}
  \frac{1}{(m_{\text{eff}}'')^{3}} = \sum_i \left(\frac{\rho_{h, i}}{\rho_{h,\text{ tot}}}\right)^2\frac{1}{m_i^{3}}.  
\end{equation}
This expression is consistent with \cite{2301.07114}: for $N$ identical copies and contributing equally to the halo density, the heating rate scales as $\propto 1/(Nm^3)$, while for $N$ copies of different mass with equal density contributions, the lightest copy dominates the heating rate, which is now $\propto 1/(N^2m_L^3)$. Extending this further, we argue that $m_{\text{eff}}''$ (like $m_{\text{eff}}'$) will in general exhibit radial variation in the multi-copy context, likely with an overall radial decrease as heavier copies may be more centrally concentrated \citep{2309.05694}. As a consequence, the disc-thickening effect may be weaker and less visible near the centre, although there, quasi-particle approximation would break down, and heating by soliton random walk would dominate more strongly \citep{Andrew2026PhRvD}. To fully assess the impact of this radial variation in $m_{\text{eff}}''$ on stellar heating, improved modelling of multi-copy halo density profiles is needed, as well as stellar scattering simulations analogous to those of \citet{Chow, 2308.14664}, but carried out specifically for the multi-copy case.

\section{Summary}
\label{sec_summary}
We have explored the transfer function for a multi-copy axion universe, providing a simplified analytic expression that can be used for future simulations. In particular, we showed that multi-copy transfer functions can be parametrised as an interpolating sum of single-copy axion transfer functions, which are easier to compute (e.g. via axionCAMB). This parametrisation can be straightforwardly extended to any number of axion copies with arbitrary relative density contributions. While we did not provide the full numerical integration of multi-copy power spectrum, our interpolating sum ansatz was tested against the 2FDM integrator of \cite{Luu2024} and reproduced the 2FDM matter power spectrum well with a $\lesssim 0.063\%$ difference in $\sigma^2(M)$. The 2FDM integrator can be used and extended to construct a multi-copy numerical integrator.
 
We also identified an equivalence among all axion copies with respect to the total density contrast $\delta_{\text{tot}}$ on scales larger than the effective Jeans scale (set by $\tilde{k}$), arising from their mutual coupling to the gravitational potential, which in turn is a manifestation of a potential gravitational equilibrium. This equivalence was previously noted by \citet{2303.00999}, but we removed their assumption that particle dark matter dominates the CDM and considered a purely $\psi$DM-dominated universe with an arbitrary number of axion copies. The equivalence implies that the large-scale structure formed in a multi-copy $\psi$DM universe might resemble an equivalent single-copy universe, with the effective mass of the equivalent axion given by
\begin{equation}
    \frac{1}{m_{\text{eff}}^2}  = \sum_i \frac{\overline{\rho_i}}{\overline{\rho_{\text{tot}}} } \frac{1}{m_i^2}.
\end{equation}
Accordingly, the onset of the faint-end turnover in galactic luminosity functions is expected to be determined solely by this effective mass scale. 

On smaller galactic scales, effects of independent axion copies become apparent through the decoupling discussed above, leading to different galaxy groups dominated by different copies, as recently reported in the simulation of \citet{Pozo2025}. This decoupling at small scales and equivalence on large scales might then collectively explain the preference for an ultra-light $\psi$DM with mass $10^{-22}$ eV from local galaxy-scale observation despite stringent mass constraints from the early Universe. The precise determination of the halo-mass function and galactic luminosity function, however, remains best investigated with improved higher-resolution cosmological simulations.

Finally, we note that for gravitational lensing (and similarly, for stellar heating with $m_{\text{eff}}''$), the net perturbation of the lensing potential induced by wave interference can be described by a separate effective scale $m_{\text{eff}}'$ that can vary locally due to the spatially dependent relative contributions of the different axion copies to the projected surface density. The detection of such a spatial variation within a single halo, for example via the spread of transients along lensed arcs from sources at different redshifts using future JWST observations, would provide direct evidence of multiple axion copies.

\begin{acknowledgements}
Authors thank referee and Ra\'{u}l Angulo for providing useful comments on improving the manuscript. Authors also thank Hoang Nhan Luu and Jennifer Fabà-Moreno for helpful discussion and sharing their 2FDM power spectrum. J.Z. and T.B. acknowledge the CEX2024-001491-S grant, funded by MICIU/AEI/10.13039/501100011033. J.Z., J.L. and S.K.L. acknowledge RGC/GRF 17312122 issued by the Research Grants Council of Hong Kong SAR. TB and PM are supported by the Spanish Grant PID2023-149016NB-I00 (MINECO/AEI/FEDER, UE). T.B., J.L. and S.K.L. acknowledge also the Collaborative Research Fund under grant C6017-20G, which is issued by the Research Grants Council of Hong Kong SAR. PM acknowledges also financial support from fellowship PIF22/177 (UPV/EHU).
\end{acknowledgements}

\bibliographystyle{aa} 
\bibliography{reference}

\end{document}